\documentclass{eas}

\usepackage{graphicx}

\begin{document}

\TitreGlobal{Mass Profiles and Shapes of Cosmological Structures}




\title{A Primer to Relativistic MOND Theory}

\author{Jacob D. Bekenstein}\address{Racah Institute of Physics, Hebrew University of Jerusalem, Jerusalem 91904 Israel}

\author{Robert H. Sanders}\address{Kapteyn Astronomical Institute, P.O. Box 800, 9700AV Groningen, The Netherlands}

\runningtitle{Relativistic MOND Theory}

\setcounter{page}{23}


\index{Jacob D. Bekenstein}

\index{Robert H. Sanders}


\begin{abstract} 
We first review the  nonrelativistic lagrangian theory as a framework for the MOND equation.  Obstructions to a relativistic version of it are discussed leading up to T$e$V$e$S, a relativistic tensor-vector-scalar field theory which displays both MOND and Newtonian limits.  The whys for its particular structure are discussed and its achievements so far are summarized.

 \end{abstract}

\maketitle

%

%




\section{Introduction}
\label{intro}

The success of Milgrom's MOND paradigm (Milgrom 1983)  in explaining the shapes of disk galaxy rotation curves, and correlations between the parameters of such systems (e.g. Tully-Fisher law) with no appeal to dark haloes is by now well known---for a modern review see Sanders \& McGaugh (2003; henceforth S\&McG).  This nonrelativistic scheme is summed up by a novel relation between acceleration ${\bf a}$ of each element of an extragalactic system and the local Newtonian field $-\nabla\Phi_N$,
\begin{equation}
\mu(|{\bf a}|/a_0){\bf a}= - \nabla\Phi_N,
\label{MOND}
\end{equation}
where $a_0\approx 10^{-10} {\rm m\ s^{-2}}$, and the $\mu$ function interpolates between $\mu(x)=x$ for $x\ll 1$ and $\mu(\infty)=1$.  It is this last proviso which guarantees correspondence of MOND with Newtonian gravity in the laboratory or the solar system, both of which are supra $a_0$ situations.

MOND lends itself to reformulation as a gravitational field theory alternative to Newton's.  In this lagrangian based theory, named AQUAL (Bekenstein \& Milgrom 1984; henceforth B\&M), the Poisson equation is replaced by
\begin{equation}
{\nabla}\cdot\left[\mu(|{\nabla}\Phi|/{a}_0){\nabla}\Phi\right]=4\pi G\rho,
\label{AQUAL}
\end{equation}
with $\Phi$ the physical gravitational potential (so that ${\bf a}=-\nabla\Phi$).   In spherically symmetric situations Eq.~(\ref{AQUAL}) has Eq.~(\ref{MOND}) as a unique solution.  But when symmetry is lower,
an extra term appears on the r.h.s.   It is this corrected MOND equation which is to be used in place of Eq.~(\ref{MOND}): deriving as it does from a lagrangian, it guarantees that momentum, energy and angular momentum will be conserved in MOND dynamics.  The original MOND formula failed to assure this.  AQUAL has a number of further advantages over plain MOND (B\&M, S\&McG). 
 
AQUAL is thus a fruitful tool for exploring non dark matter (DM) resolutions of the missing mass problems.  But it has shortcomings.  For example, like the naive MOND formula, it fails to do away with the need for DM in clusters of galaxies, although its requirements for the dark stuff are certainly lower than in Newtonian gravity.  And since it is basically nonrelativistic, AQUAL is unable to address a number of key subjects with relativistic overtones: gravitational lensing, cosmological evolution and the growth of structure, to name a few.  How to cast the MOND idea in a relativistic mold while retaining the benefits of AQUAL ?

\section{Relativistic MOND theories leading up to  \textrm{T$e$V$e$S}}
\label{relativistic}

A relativistic version of AQUAL (call it RAQUAL) was offered in its defining paper (B\&M).  The basic idea was to replace the physical potential $\Phi$ by a scalar field $\phi$ whose equation is a relativistic reflection of Eq.~(\ref{AQUAL}).  This equation is written on a metric $g_{\alpha\beta}$  obeying Einstein's equations (with a contribution to the energy momentum tensor from $\phi$).  One then defines the physical metric as
\begin{equation}
\tilde g_{\alpha\beta}\equiv e^{2\phi} g_{\alpha\beta}.
\label{conf}
\end{equation}
This last describes a spacetime which has been locally stretched in every direction with respect to that described by $g_{\alpha\beta}$.  All matter and radiations are supposed to sense only $\tilde g_{\alpha\beta}$, and the requisite equations must contain only $\tilde g_{\alpha\beta}$ and never $g_{\alpha\beta}$ by itself.  RAQUAL reproduces nonrelativistic AQUAL, and so recovers much of the successful MOND phenomenology.  

However, RAQUAL fails on two counts.  It permits superluminal propagation of $\phi$ waves (B\&M).   And it is unable to give an account of gravitational lensing in agreement with the basic observation that lensing by galaxy clusters is anomalously strong compared to what was to be expected in view of their galaxies and gas content.  This last problem is rooted in the conformal relation (\ref{conf}) between the Einstein and physical metrics.  Conformally related metrics share the same tracks for photons (or electromagnetic wave packets).  And because $\phi$'s contribution to the source of Einstein's equations is weak in clusters of galaxies, the metric relevant for lensing in RAQUAL is nearly  that in GR with no DM.  

Breaking the conformal relation has been the central theme in attempts to fix this problem.  For example, it is possible to add to Eq.~(\ref{conf}) the symmetric tensor $\phi_{,\alpha}\,\phi_{,\beta}$ multiplied by some scalar (Bekenstein \& Sanders 1994).  It turns out that in order for gravitational waves---perturbations of $g_{\alpha\beta}$---to propagate causally, i.e. within the light cone of $\tilde g_{\alpha\beta}$, the mentioned scalar factor must be negative, and it is then found to \emph{weaken} gravitational lensing, rather than enhancing it as intended.  

A way out of the \emph{impasse}  (Sanders 1997) is to introduce a constant vector field which points in the time direction, and then to stretch spacetime along the vector by a factor $e^{2\phi}$ while shrinking it by the same factor orthogonally to the vector.  This does the trick: provided the scalar field $\phi$ used comes from a RAQUAL-type equation, MOND phenomenology is recovered while the lensing is augmented to the right proportions.  However, this ``Stratified'' theory is not covariant (one needs the vector to be in the time direction, and this cannot be true in all coordinate systems).  Covariance of equations of a physical theory is so ingrained in modern physics that we have to look further.

\section{The \textrm{T$e$V$e$S} field theory}
\label{TeVeS}

\textrm{T$e$V$e$S} is a tensor-vector-scalar covariant field theory (Bekenstein 2004; henceforth B04) incorporating various motifs already mentioned.  Just as in RAQUAL, the tensor is an Einstein metric $g_{\alpha\beta}$ out of which is built the standard Einstein-Hilbert action.  One passes to the physical metric as in the stratified theory, but now using an everywhere timelike \emph{dynamical} normalized vector field ${\cal U}_\alpha$.  In equations
\begin{equation}
\tilde g_{\alpha\beta}=e^{-2\phi}(g_{\alpha\beta} + {\cal U}_\alpha {\cal U}_\beta) - e^{2\phi} {\cal U}_\alpha {\cal U}_\beta,
\label{metric}
\end{equation}
together with the normalization requirement $g^{\alpha\beta}{\cal U}_\alpha{\cal U}_\beta=-1$.  

The dynamics of ${\cal U}_\alpha$ is dictated by the action
\begin{equation} S_v =-{K\over 32\pi G}\int
\big[g^{\alpha\beta}g^{\mu\nu} 
{\cal U}_{[\alpha,\mu]} {\cal U}_{[\beta,\nu]}
-2(\lambda/K)(g^{\mu\nu}{\cal U}_\mu
{\cal U}_\nu +1)\big](-g)^{1/2} d^4 x,
\label{vector_action}
\end{equation}
where the square brackets denote antisymmetrization. The $K$ here is a \emph{dimensionless} coupling constant, one of the parameters of the theory.  The $\lambda$ is a Lagrange multiplier field included to enforce the normalization of the vector field; $\lambda$ is to be determined as the equations are solved.  The kinetic part of $S_v$ is that of an Abelian gauge field, but the unit norm constraint deprives ${\cal U}_\alpha$ of the proverbial gauge freedom.   In addition, ${\cal U}_\alpha$ is not coupled to a current of any sort, so it would be incorrect to view the vector as mediating a repulsive Coulomb-like force.

The dynamical scalar field $\phi$ required in Eq.~(\ref{metric}) is assigned the action
\begin{equation} S_s =-{\scriptstyle 1\over\scriptstyle
2}\int\big[\sigma^2
h^{\alpha\beta}\phi_{,\alpha}\phi_{,\beta}+{\scriptstyle
1\over\scriptstyle 2}G
\ell^{-2}\sigma^4 F(kG\sigma^2) \big](-g)^{1/2} d^4 x,
\label{scalar}
\end{equation} 
which involves an auxiliary  \emph{nondynamical} scalar field $\sigma$, a new dimensionless parameter, $k$,  a scale parameter, $\ell$, and an unspecified function $F$.    In the kinetic part of $S_s$,  $\phi_{,\alpha}\phi_{,\beta}$ is contracted not with $g^{\alpha\beta}$, but with the tensor $h^{\alpha\beta}\equiv g^{\alpha\beta} -{\cal U}^\alpha {\cal U}^\beta$; this is introduced to cope with the problem of superluminal propagation of scalar perturbations evidenced by RAQUAL.  With $h^{\alpha\beta}$ the propagation is found to be subluminal with respect to the physical metric on condition that $\phi>0$ (B04).   Contrary to appearances, $S_s$ is not a quadratic action.  First $\sigma^2$ must  be expressed in terms of $\phi_{,\alpha}$ by varying $S_s$ with respect to $\sigma^2$, and inverting the resulting equation.  When the result is substituted in both terms of $S_s$, the action becomes aquadratic in $\phi_{,\alpha}$, resembling RAQUAL's action, with the scalar $h^{\alpha\beta}\phi_{,\alpha}\,\phi_{,\beta}$ replacing $g^{\alpha\beta}\phi_{,\alpha}\,\phi_{,\beta}$.  The form (\ref{scalar}) is, however, more convenient for calculation.

These pieces of the total action must be supplemented with the matter action $S_m$.  In accordance with the Einstein equivalence principle (Will 1993) it must be formed by replacing $g_{\alpha\beta}\mapsto\tilde g_{\alpha\beta}$ everywhere in the matter action in use in general relativity (GR), and modifying all covariant derivatives accordingly.   In this way all types of matter respond to gravitational fields in the same way, and it could be said that matter and radiation delineate the physical metric.  Now because $\tilde g_{\alpha\beta}$ contains $\phi$ as well as ${\cal U}_\alpha$, the variations of the total action with respect to these quantities engender variations of $S_m$ which create sources for the equations of $\phi$ and ${\cal U}_\alpha$.  No direct coupling of these fields to the matter variables is thus required, which is pleasant as no principles are evident for determining such couplings.

Is T$e$V$e$S unique ?  Not at all.  For example, one could choose to write $S_s$ and $S_v$ with $\tilde g_{\alpha\beta}$ instead of $g_{\alpha\beta}$.  The present approach seems preferable in that evidently $g_{\alpha\beta}$, $\phi$ and ${\cal U}_\alpha$ are all gravitational fields, so their actions should not involve the ``matter's metric''  $\tilde g_{\alpha\beta}$.  Further, one could add to $S_s$ a kinetic term for $\sigma$ such as is included in Phase Coupled Gravity (PCG), a theory devised in a partially successful attempt to forestall superluminal propagation (Bekenstein 1986, 1990).    In addition one could redefine $h^{\alpha\beta}$ as $g^{\alpha\beta}-\zeta\,  {\cal U}^\alpha\,  {\cal U}^\beta$ ($\zeta$ should be positive but not small in order to obviate superluminality; it could be a constant).   An additional freedom is the choice of function $F$ in Eq.~(\ref{scalar}); each such choice defines a distinct theory.   Some of the freedoms mentioned have been  used to make a variant of T$e$V$e$S (Sanders 2005a, 2005b; henceforth S05).

\section{Results from \textrm{T$e$V$e$S}}
\label{results}

T$e$V$e$S has GR as a limit for $K\to 0$ and $\ell\to \infty$ (B04).  Thus if $|K|\ll 1$ and the size of the system of interest falls well short of $\ell$, predictions of T$e$V$e$S and GR should not be that different.   To verify that T$e$V$e$S can also give MOND in some other limit one must make a choice for the function $F$ in action (\ref{scalar}).  B04 makes the choice which is equivalent to the relation
\begin{equation} 
k\ell^2 h^{\mu\nu}\phi_{,\mu}\phi_{,\nu}=(3/4) k^2 G^2 \sigma^4(k G \sigma^2-2)^2 (1-kG\sigma^2)^{-1}.
\label{y}
\end{equation}
This is to be regarded as defining a toy theory.  A better choice would have the r.h.s. change sign smoothly through $kG\sigma^2=1$, a transition corresponding to the passage between a quasistatic system such as a cluster of galaxies ($h^{\mu\nu}\phi_{,\mu}\phi_{,\nu}>0$) and cosmology ($h^{\mu\nu}\phi_{,\mu}\phi_{,\nu}<0$) (see B04).

Imagine linearizing the Einstein-like equations of T$e$V$e$S about flat spacetime ($g_{\alpha\beta}=\eta_{\alpha\beta} + \cdots $) in a quasistatic situation.  The final result for the physical metric after suitable choice of units is (B04)
\begin{equation}ds^2
=-(1+2\Phi)dt^2+(1-2\Phi)\delta_{ij} dx^i dx^j;\qquad \Phi\equiv \Xi\Phi_N+\phi.
\label{weakmetric}
\end{equation}
Here $\Phi_N$ is the usual Newtonian potential of the baryonic matter and $\Xi$ is a constant depending on $K$ and on the asymptotic (cosmological) value of the scalar field $\phi$; $\Xi$ is expected to be close to unity.  With $\Phi$ interpreted as the physical gravitational potential, metric (\ref{weakmetric}) is precisely that used in GR to discuss both dynamics \emph{and} gravitational lensing. An immediate conclusion is that if GR succeeds in describing both the dynamics of, say, a galaxy, and its lensing properties with a \emph{single} assumed DM distribution, and if T$e$V$e$S succeeds in predicting the dynamics everywhere from the observed baryonic distribution, then T$e$V$e$S will predict the same lensing pattern as GR.  Thus the problem that accosted RAQUAL and variants of it (Sec.~\ref{relativistic}) is solved: lensing can be described using MOND ideas.  This much was clear earlier from the non-covariant framework (Sanders 1997).  Of course the problem with galaxy cluster dynamics (Sec.~\ref{intro}) still looms. But it is important to realize that the two mentioned problems are distinct.

Obviously, to determine $\Phi$ requires one to calculate $\phi$ from the baryonic mass distribution using the T$e$V$e$S analog of Eq.~(\ref{AQUAL}).  In the linearized framework mentioned and assuming spherical symmetry, one can be immediately integrate the said equation once to get (we assume $k\ll 1$ as well as $\Xi\approx 1$)
\begin{equation}
\mu \nabla\Phi = \nabla\Phi_N, \qquad \mu\approx {-1+\sqrt{1+4|\nabla\Phi|/
{a}_0} \over 1+\sqrt{1+4|\nabla\Phi|/ {a}_0} };\qquad {a}_0\approx  {\sqrt{3 k}\over 4\pi\ell} .
\label{mu}
\end{equation}
Thus we get the MOND equation (\ref{MOND}), as well as a formula for $\mu$ showing the appropriate asymptotic behavior at small and large $|\nabla\Phi|/ {a}_0$, and the value of $a_0$ expressed in terms of the T$e$V$e$S parameters.  MOND is recovered.   Famaey and Binney (2005) have noted that the Galaxy's rotation curve is not fully consistent with this $\mu$.  Apparently a better $F$ is needed.

\begin{figure}[h]
\centering
\includegraphics[width=6cm]{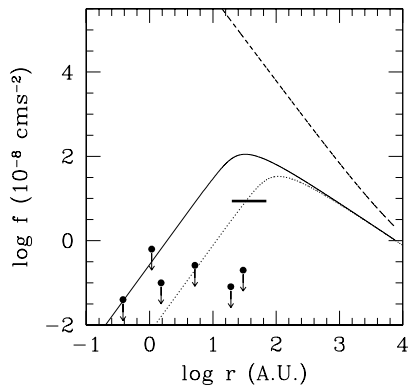}
  \caption{$|\nabla\Phi|$ (dashed),  $|\nabla(\Phi-\Phi_N)|$
 (in units of $10^{-8}$ cms$^{-2}$) with $k=0.03$ 
(solid)  or $k=0.1$ (dotted) from T$e$V$e$S plotted vs the distance from 
the Sun.  Observed constraints on the non-Newtonian part of the 
acceleration, $|\nabla(\Phi-\Phi_N)|$, are (left to right): 
from the precession of perihelion of Mercury, and of Icarus, 
from variation of Kepler's constant between Earth and Mars, between 
inner planets and Jupiter,  Uranus or Neptune, respectively.  
The solid bar is the Pioneer anomaly range. }
 \label{figure:fig1}
 \end{figure}

The approximation inherent in Eq.~(\ref{mu}) is good for all extragalactic applications, but begins to fail when $|\nabla\Phi|> (\pi/k)^2 a_0$.  In such situations, e.g. the solar system,  it is best to start from the first term of the Laurent  expansion of the r.h.s. of Eq.~(\ref{y}) about $kG\sigma^2=1$.  One then obtains
\begin{equation}
\mu\approx  {1\over \Xi+k/4\pi}
\left(1-{16\pi^3\over  k^3}{{a}_0{}^2\over
|\nabla\Phi|^2}\right).
\label{mu2}
\end{equation} 
The fact that $\mu$ asymptotes to $(\Xi+k/4\pi)^{-1}$ as $|\nabla\Phi|\to \infty$ means that for arbitrarily strong fields gravity is Newtonian, except that the  effective gravitational constant is  $G_N=(\Xi+k/4\pi)\,G$.  It should be mentioned that the salient post-Newtonian parameters (Will 1993), important for inner solar system celestial dynamics, agree with those of GR; in particular $\beta=1$ and $\gamma=1$ (Giannios 2004, B04).

Notice the slow approach of $\mu$ to its asymptotic value as $|\nabla\Phi|$ strengthens. One consequence is a slow increase of the Kepler constant $a^3/P^2$ with distance from the Sun.  Fig. 1 illustrates this effect and compares it with the constraints set by measurements of the orbits of diverse planets.   Although a better form of $F$ is clearly required, the significant point is that, in such theories, the total force in the outer Solar System will inevitably deviate from inverse square.  This is of considerable interest in view of the Pioneer anomaly (Anderson et al. 1998,
Turyshev 2005).


\begin{thebibliography}{}


\bibitem {} Anderson, J.D. et al., 1998, Phys.Rev.Lett., 81, 2858

\bibitem{}Bekenstein, J.D. \& Milgrom, M. 1984, ApJ, 286, 7 (B\&M)

\bibitem{}Bekenstein, J.D. 1988, in Second Canadian Conference  on General Relativity and Relativistic Astrophysics, eds. A. Coley, C. Dyer \& T. Tupper (Singapore: World Scientific), 68. 

\bibitem{}Bekenstein, J.D. 1990, in Developments in General Relativity, Astrophysics and Quantum Theory, eds. F. I. Cooperstock, L. P. Horwitz \& J. Rosen (Bristol: IOP Publishing), 156

\bibitem{}Bekenstein, J. D. \& Sanders, R.H. 1994, ApJ, 429, 480

\bibitem{}Bekenstein, J.D. 2004, Phys. Rev. D, 70, 083509 (erratum, 2005, Phys. Rev. D, 71, 069901) (B04)

\bibitem{}Famaey, B. \& Binney, J.  2005, astro-ph/0506723

\bibitem{}Giannios, D. 2005 Phys. Rev.  D, 71, 103511

\bibitem{}Milgrom, M. 1983, ApJ, 270, 365, 379 and 384 

\bibitem{}Sanders, R.H.1997, ApJ, 480, 492

\bibitem{}Sanders R.H. \& McGaugh, S.S.  2002, Ann.
Rev. Astron. Astrophys., 40,  263 (S\&McG)

\bibitem{}Sanders, R.H. 2005a, astro-ph/0502222

\bibitem{}Sanders, R.H. 2005b, this collection (S05)

\bibitem{} Turyshev, S., 2005, this collection

\bibitem{}Will, C. 1993, Theory and Experiment in Gravitational Physics (Cambridge: Cambridge Univ. Press)


\end{thebibliography}
\end{document}